\documentclass[11pt,oneside,letterpaper]{article}
\usepackage{amssymb}
\usepackage{amsmath}
\usepackage{graphicx}
\usepackage{setspace}
\usepackage{fancyhdr}
\usepackage{ifpdf}
\usepackage{graphicx}
\usepackage{comment}
\def\ip{${\mathcal{I}^+}$}
\def\half{{\textstyle{1\over2}}}

 \def\p{\partial}

 \def\dsp{$dS^+$}

\newcommand{\bea}{\begin{eqnarray}}
\newcommand{\eea}{\end{eqnarray}}
\newcommand{\be}{\begin{equation}}
\newcommand{\ee}{\end{equation}}

\newcommand{\bi}{\begin{itemize}}
\newcommand{\ei}{\end{itemize}}

\numberwithin{equation}{section}
\allowdisplaybreaks  

\begin{document}

\begin{center}
{ \LARGE \textsc{Asymptotic Symmetries and Charges in De~Sitter Space}}
\begin{center}
Dionysios Anninos, Gim Seng Ng and Andrew Strominger
\end{center}
\end{center}
\vspace*{0.6cm}
\begin{center}
Center for the Fundamental Laws of Nature, \\ Harvard University, Cambridge, MA 02138, USA
\vspace*{0.8cm}
\end{center}
\vspace*{1.5cm}
\begin{abstract}
\noindent
{
\newline
}
\end{abstract}
The asymptotic symmetry group (ASG) at future null infinity (\ip ) of four-dimensional de Sitter spacetimes is defined and shown to be given by the group of three-dimensional diffeomorphisms acting on \ip.  Finite charges are constructed for each choice of ASG generator together with a  two-surface on \ip. A conservation equation is derived relating the evolution of the charges with the radiation flux through \ip.

\newpage
\onehalfspacing
\tableofcontents
\section{Introduction}

Diffeomorphisms are the basic symmetry of general relativity. In spacetimes with an asymptotic boundary, there is an interesting subgroup of the diffeomorphisms, often referred to as the asymptotic symmetry group (ASG), which ``acts nontrivially" on the boundary data. The precise form of the ASG depends on the spacetime in question, the boundary conditions, the dynamics and the precise definition of ``acts nontrivially". In some cases charges can be associated with the asymptotic  symmetries.

The subject of asymptotic symmetries began with the seminal work of Arnowitt, Deser and Misner \cite{Arnowitt:1959ah,Arnowitt:1962hi} who showed that the ADM energy and momentum of an asymptotically Minkowski spacetime are associated with asymptotic translations at spatial infinity.  Bondi, Meissner and Sachs \cite{Bondi:1962px,Sachs:1962wk,Sachs:1962zza,Penrose:1962ij} studied the asymptotic symmetries at future null infinity (\ip) and discovered an infinite-dimensional group known as the BMS group. The structure at \ip\ is much more complicated than the one at spatial infinity because radiation can pass through \ip. More recently \cite{Geroch80,Wald:1999wa}, integrable charges have been constructed which are parameterized by both a choice of generator of the BMS group and a two-surface on \ip. These charges obey a conservation equation reflecting the possibility of radiation flux at \ip \cite{Ashtekar:1981bq,dray}.


In this paper we give a definition of  the ASG for four-dimensional spacetimes, denoted \dsp, which are asymptotically de Sitter. Such spacetimes are of special interest because they may include the one we inhabit. The asymptotic region \ip\ of such spacetimes is a spacelike surface which coincides with future timelike infinity. Its structure is in some ways similar to Minkowskian \ip\ because there is in general a nonzero radiation flux, and an infinite-dimensional group is expected. We show that the ASG so defined is all of the three-dimensional diffeomorhpisms acting on \ip.\footnote{Although the considerations of this paper are purely classical, we note in passing that the ASG defined here is not necessarily  a candidate symmetry group for a holographic dual for \dsp\  quantum gravity as the latter may involve a different treatment of \ip.}

A naive power-counting of falloffs of the relevant structures as \ip\ as approached indicates that the charges associated with the ASG for \dsp\ should be divergent and depend
unpleasantly on the manner in which \ip\ as approached. However adapting the results and insights of a number of recent papers \cite{Wald:1999wa,Barnich:2001jy,Compere:2008us, Anninos:2009yc}, we find that finite charges which do not depend on the approach to \ip\ can in fact be constructed. The resulting expression for the charges (\ref{wz}) and their conservation equation (\ref{fluxI}) below are our  main results.

In judicious circumstances charges of the type constructed here generate, via Dirac brackets, the associated symmetries. This may be the case for our charges , but subtleties of the type discussed in \cite{Ashtekar:1981bq} for the BMS charges (see also \cite{Hollands:2005ya}) must be addressed to demonstrate this.

This paper is organized as follows. In section \ref{bcs} we discuss the asymptotic expansion of $dS^+$ spacetimes near future infinity. In section \ref{ASG} we propose our definition for the asymptotic symmetry group of $dS^+$. Finite charges for those diffeomorphims tangent to $\mathcal{I}^+$ comprising the ASG are constructed in section \ref{BY}. These charges are refined using covariant phase space techniques in section \ref{cov}. Two appendices providing details of the calculations are given at the end. 

\section{Boundary conditions}\label{bcs}

We are interested in solutions of the Einstein equations with a positive cosmological constant $\Lambda={3 / \ell^2}$:\footnote{We do not include addtional matter in our discussions.  It may not be difficult to do so, especially if it is assumed that everything ultimately decays into massless particles. }
\be\label{xwq}
R_{\mu\nu} - \frac{1}{2}R g_{\mu\nu} = - {3 \over \ell^2} g_{\mu\nu}~.
\ee de Sitter space is the simplest solution of (\ref{xwq}).
More complicated solutions often have singularities either in the future or the past.  In this paper we are mainly interested in solutions which are asymptotically future de Sitter, which we will refer to as \dsp\ spacetimes. That is, they may be singular in the past, but in some sense approach pure de Sitter in the future. \dsp spacetimes are of special interest as they may well include our own universe!

In order to define our notion of a \dsp  spacetime we must specify boundary conditions. We wish to make the boundary conditions relatively tight, to simplify the calculations, while still allowing for example the possibility of gravitational waves to reach \ip.  It was shown by  Starobinsky \cite{Starobinsky:1982mr} that a very general class of excited  de Sitter spacetimes can at late times be put in the ``Fefferman-Graham" (FG) form \cite{fefferman}
\be\label{FG}
\frac{ds^2}{\ell^2} = -\frac{d\eta^2}{\eta^2} + \frac{dx^i dx^j}{\eta^2} \left(  g^{(0)}_{ij}  + \eta^2 g^{(2)}_{ij}  + \eta^3 g^{(3)}_{ij}   + \ldots \right)~.
\ee
Here \ip\ is $\eta \to 0$. Note that the term proportional to $\eta^{-1}$ can be and is set to zero in these coordinates.
For the purposes of this paper, we take the definition of a \dsp\ spacetime to be any solution of the Einstein equation (\ref{xwq})
with an expansion of the form (\ref{FG}) with the $g^{(k)}$ smooth tensors on ${\mathbb R}^3$.

The Einstein equation imposes relations
among  the coefficients in the FG  expansion. These include:
\begin{eqnarray}
g^{(2)}_{ij} &=& R_{ij}[g^{(0)}] - \frac{1}{4} R[g^{(0)}] g^{(0)}_{ij}~,\\
\nabla^j g^{(3)}_{i j} &=& \text{tr} \; g^{(3)}_{ij} = 0~,
\end{eqnarray}
where the covariant derivative and trace are defined with respect to $g^{(0)}$. Moreover the coefficients $g^{(k)}$ of $\eta^{k-2}$ for $k>3$ are then fully  determined by $g^{(0)}$ and $g^{(3)}$. Hence the data characterizing the spacetime is a boundary metric $g^{(0)}$ and a traceless conserved tensor $g^{(3)}$.

\subsection{Conformal slicing transformations}

The precise forms of the boundary data $g^{(0)}$ and $g^{(3)}$ depend on the precise choice of slices labeled by constant $\eta$ as \ip is approached. Consider an infinitesimal slicing transformation characterized by the diffeomorphism $\eta \to \eta-\eta\delta \sigma(x^i)$. In order to preserve FG gauge (\ref{FG}), this must be accompanied by an $\eta$-dependent diffeomorphism tangent to the slice. FG gauge-preserving slicing transformations are generated by the vector fields
\be
\label{slc}
 \xi^{(\delta\sigma)} = \eta \delta \sigma (x^k)\p_\eta
+ \ell^2 \left[\partial_j \delta \sigma(x^k) \right] \int^\eta  \frac{d\eta'}{\eta'} g^{ij}(\eta',x^k) \partial_i~.
\ee
These slicing transformations act as conformal transformations on the data at \ip\ according to
\begin{eqnarray}
\delta g^{(0)}_{ij} &=& 2\delta \sigma  g^{(0)}_{ij}~, \\
\delta g^{(3)}_{ij} &=& -\delta \sigma g^{(3)}_{ij}~.
\end{eqnarray}
We see that $g^{(0)}$ transforms with weight 2 while $g^{(3)}$ transforms with weight -1. Hence the physical boundary data, which do not depend on a slicing choice near \ip\ are a conformal metric of conformal weight 2 and a traceless symmetric conserved tensor of weight -1.

\section{The Asymptotic Symmetry Group}\label{ASG}

The asympotic symmetry group (ASG) is defined as the quotient group
\be \text{ASG} \equiv {\rm allowed ~diffeomorphisms \over trivial ~diffeomorphisms}~.
\ee
In the present case, an allowed diffeomorphism is any one which preserves the FG form of every \dsp\ metric. We define the trivial diffeomorphisms
to be those which leave the boundary data on \ip\ invariant.

The most general allowed diffeomorphism is of the form
\be\label{gdf}  \xi =  \phi^i(x^k)\p_i +  \eta \delta \sigma (x^k)\p_\eta+\frac{\eta^2}{2}  g^{(0)ij} \partial_j \delta \sigma(x^k)\p_i  + {\cal O}(\eta^4)~,
\ee
where here and hereafter $i,j$ indices of $g_{ij}^{(k)}$ are raised and lowered with $g^{(0)}_{ij}$.

As seen above, the $\delta \sigma$-dependent terms do not change the boundary data. Moreover the subleading $\eta$ terms fall off so rapidly that they also do not change the boundary data. Hence only the first term in (\ref{gdf}) is non-trivial, and the ASG is generated by diffeomorphisms of the form
\be \label{asg} \xi^{\text{ASG}} =  \phi^i(x^k)\p_i~.\ee
Hence the ASG is simply the diffeomorphisms of ${\mathbb R}^3$.

The short and simple treatment we have given here parallels the the one which originally lead to the BMS group as the ASG for the  Minkowski space \ip. A somewhat unsatisfactory feature of this treatment is that it is not clear precisely what is non-trivial about the remaining ${\mathbb R}^3$ diffeomorphisms. The best answer to this question is (and will be below) provided by a more elaborate discussion involving the construction of asymptotic charges transforming non-trivially under the ASG.  The trivial diffeomorphisms are defined as those
whose associated charges or generators vanish when the constraints are applied.  However when there are energy fluxes through the boundary, as is the case for \ip\ in either Minkowski or de Sitter spacetimes, the charges will not be conserved and are rather subtle to define. The definition of  such charges for Minkowski \ip\cite{Geroch80,Wald:1999wa,dray} did not appear until several decades after the original work of BMS, and requires considerably more technology.  Using
the vanishing of these charges as the definition of triviality, one indeed recovers the BMS group.

Fortunately the hard work which went in to the  definition of asymptotic charges at Minkowksi \ip\ along with the development of the covariant phase space formalism \cite{Barnich:2001jy,Iyer:1994ys} can be readily adapted to the \dsp\ case. We will see in the following that defining a trivial diffeomorphism as one whose associated charge vanishes happily leads back to the conclusion stated above that the \dsp\ ASG is all of the diffeomorphisms of $\mathbb{R}^3$.

\section{Brown-York charges}\label{BY}

In this section we will use the Brown-York formalism \cite{Brown:1992br} to compute the charges associated with the ASG generators (\ref{asg}) as well as their conservation laws.

The Brown-York formalism as developed in \cite{Brown:1992br} is relevant only for diffeomorphisms which do not move the boundary of a given hypersurface. They therefore cannot be used to associate charges or determine triviality of the more general allowed diffeomorphisms (\ref{gdf}).
This requires the more fully developed but also more complicated covariant phase space formalism, which will see in the following section in the end  leads back to the same expression for the charges found more simply in this section.

\subsection{Charges}

The Brown-York stress tensor associated to a three-dimensional hypersurface $\Sigma$, including possible counterterms required for finiteness
at $\mathcal{I}^+$ of \dsp,  is given by :
\be \label{byst}
T_{ij} = -\frac{1}{8\pi G} \left( K_{ij} - K \gamma_{ij} - c_1\frac{2}{\ell} \gamma_{ij} - c_2 \ell G_{ij}  \right)~.
\ee
Here $\gamma_{ij},~G_{ij}$ and $K_{ij}$ are the induced intrinsic metric and Einstein tensor and extrinsic curvature respectively of $\Sigma$, while $c_1$ and $c_2$ are at this point  arbitrary constants. This expression is derived  \cite{Brown:1992br,Balasubramanian:2001nb,Balasubramanian:1999re}  by taking a variational derivative of the Einstein-Hilbert action plus counterterms with respect to the induced metric at the boundary. The regularized action whose variation gives  (\ref{byst}) is:
\be\label{action}
S_{total} = \frac{1}{16\pi G} \int_\mathcal{M} d^4 x \sqrt{-g} \left( R\left[g\right] - 6/\ell^2 \right) + \int_{\mathcal{I}^+} L_{GH}  + \int_{\mathcal{I}^+} L_{ct}~.
\ee
where
\be
L_{GH} \equiv \frac{1}{8\pi G} \sqrt{\gamma} K\left[\gamma\right]d^3 x, \quad L_{ct} \equiv \frac{1}{16\pi G } \sqrt{\gamma} \left( c_2 \ell^2 R[\gamma] - 4c_1/\ell \right)d^3x~.
\ee

Evaluating the on-shell Brown-York stress tensor at a surface of small constant $\eta$ near \ip\ of an arbitrary \dsp\ spacetime one finds (along the lines of \cite{de Haro:2000xn,Henningson:1998gx,Skenderis:2002wp}):
\begin{eqnarray}
T_{ij}   & =& -\frac{\ell}{8\pi G } \left[ \frac{2}{\eta^2} \left(1-c_1\right)g^{(0)}_{ij}
+\left(3-2 c_1 \right) g^{(2)}_{ij}-\left(g^{(0)kl}g^{(2)}_{kl}\right) g^{(0)}_{ij} -c_2 G_{ij}[g^{(0)}] \right.\nonumber\\
& &\left.+\eta \left(\frac{7}{2}-2 c_1 \right) g^{(3)}_{ij}\right] +{\cal O}(\eta^2)~. 
\end{eqnarray}
In the hope of constructing finite charges at \dsp, we choose $c_1$ and $c_2$ to make this expression as small as possible for $\eta \to 0$. With the choice of \be c_1=c_2=1~, \ee we arrive at
\be
T_{ij} = -  \frac{3 \eta \ell}{16\pi G} g^{(3)}_{ij} \equiv \eta T_{(0)ij}~, \quad T^i_i = \nabla^i T_{ij} = 0~,
\ee  where we have defined the ${\cal O}(1)$ part $T_{(0)ij}$ for convenience.
Charges are associated with a two-dimensional  compact submanifold  $\p \Sigma$ of \ip, which is a boundary of a noncompact three-volume $\Sigma$, together with a vector field $\xi$ of the form (\ref{asg}) generating the ASG. We define these charges by working on a hypersurface of small constant $\eta$ and using a submanifold $\p \Sigma_\eta $ of $\Sigma$ that approaches the desired $\p \Sigma$  on \ip\ for $\eta \to 0$ and is the boundary of a noncompact  hypersurface $\Sigma_\eta$ in \dsp.  The expression is
\be\label{charge}
Q^{BY}_\xi\left[ g^{(0)}, g^{(3)}, \p \Sigma \right] = \lim_{\eta \to 0} \int_{\partial \Sigma_\eta} d^2x \sqrt{\sigma} n^i \xi^j T_{ij}~.
\ee
$n^i$  here is tangent to $\Sigma_\eta$  and normal to $\partial \Sigma_\eta$, while $\sigma $ is the induced metric on $\p\Sigma_\eta$. Finiteness of the charges follows from the fact that $n^i \sim \eta$,  $\sqrt{\sigma} \sim \eta^{-2}$ and $T_{ij}\sim \eta$. This expression is manifestly conformally invariant
\be
Q^{BY}_\xi \left[g^{(0)}_{ij},g^{(3)}_{ij}\right] = Q^{BY}_\xi \left[\Omega^2 g^{(0)}_{ij}, \Omega^{-1} g^{(3)}_{ij} \right]~.
\ee
This is equivalent to the statement that the charges are independent of the precise manner in which the surface $\p\Sigma_\eta$ is taken to the boundary \dsp. When there are appropriate asymptotic Killing vectors these Brown-York charges reduce to the conserved Abbott-Deser charges \cite{Abbott:1981ff}.
\subsection{Conservation equation}
\begin{figure}
\begin{centering}
\includegraphics[scale=0.4]{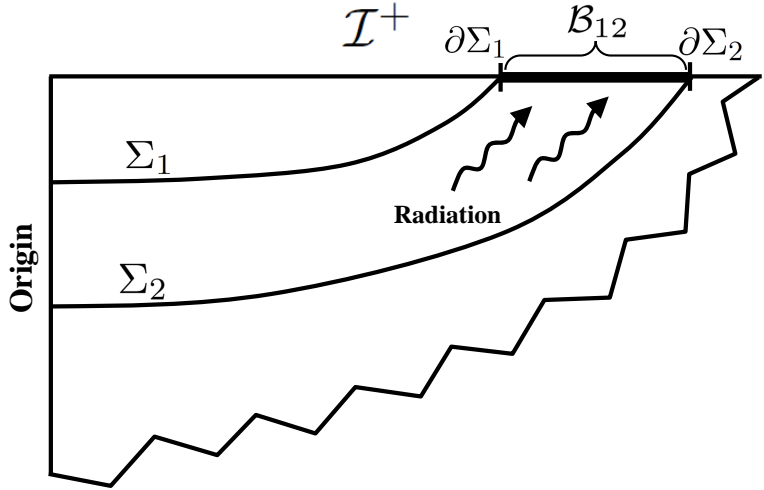}
\label{fig:flux}
\caption{ Consider two spacelike slices $\Sigma_1$ and $\Sigma_2$ ending on $\partial \Sigma_1$ and $\partial \Sigma_2$. The difference between the Brown-York charge $\Delta Q_{BY}$ is given by the integral over the radiation flux ${F}_\xi$ in the 3-volume ($\mathcal{B}_{12}$ in ${\cal I}^+$) bounded by $\partial \Sigma_1$ and $\partial \Sigma_2$. Here, the Penrose diagram depicts a spacetime which tends to de Sitter space in the far future. The spacelike jagged line represents the far past which could for example be the big bang.}
\end{centering}
\end{figure}

Consider two submanifolds  $\partial \Sigma_1$ and $\partial \Sigma_2$ of $\mathcal{I}^+$ which bound two spacelike hypersurfaces $\Sigma_1$ and $\Sigma_2$.   In general we do not expect that the charge associated to a generic vector field $\xi$ will be the same for $\partial \Sigma_1$ and $\partial \Sigma_2$, in part because energy and other fluxes can leak out through the region $\mathcal{B}_{12}$ of \ip\ between them (see Fig.~1).  Rather, we expect the difference in the charges to be related to a flux through this region. The formula for this is straightforward to obtain:
\bea\label{fluxI}
\Delta Q_{BY} &\equiv &\int_{\partial \Sigma_2} d^2x \sqrt{\sigma} n^i \xi^j T_{ij}-
\int_{\partial \Sigma_1} d^2x \sqrt{\sigma} n^i \xi^j T_{ij} \nonumber\\ &=& -\frac{3\ell}{32\pi G} \int_{\mathcal{B}_{12}}   g^{(3)ij}\mathcal{L}_\xi g^{(0)}_{ij} \sqrt{g^{(0)}} d^3x \nonumber\\ &\equiv&  \int_{\mathcal{B}_{12}} {F}_\xi~,
\eea
where we have integrated by parts and  used $\nabla^j T_{ij} = 0$.

When $\xi$ is an isometry of $g^{(0)}$, the flux vanishes.  The form of (\ref{fluxI}) is reminiscent of the analogous case in Minkowski  space where ${ F}_\xi = -\frac{1}{32 \pi G} N_{ab} \chi^{ab} {\boldsymbol{\epsilon}}^{(3)} $ with $\chi_{ab} \propto {\cal L}_\xi g_{ab}$ and $N_{ab}$ the Bondi news tensor \cite{Geroch80,Wald:1999wa,Hollands05}. In the BMS case, the flux vanishes whenever the news tensor is zero, which is equivalent to the absence of radiation. In our context, we can interpret $ g^{(3)}_{ij}$ as the `de Sitter news tensor'. When $g^{(3)}_{ij}=0$, as in pure de Sitter,  there is no radiation and no flux. Indeed, adding a gravitational wave to pure de Sitter \cite{Mukhanov:1990me} gives rise to a non-zero $g^{(3)}_{ij}$ term which is recorded by the flux.


\section{Covariant phase space charges}\label{cov}

In this section we consider  the charges given by the covariant phase space formalism \cite{Wald:1999wa,Barnich:2001jy,Iyer:1994ys}.  One advantage of these charges is that, in contrast to the Brown-York charges, they may be evaluated for diffeomorphisms that are not tangent to $\mathcal{I}^+$.\footnote{In what follows, we evaluate all expressions on-shell.}

\subsection{The symplectic form}\label{symform}

In this section we construct a symplectic form $\omega\left[\delta_1 g, \delta_2g\right]$ for the phase space of initial data on a hypersurface $\Sigma$ ending on \ip.  The problem is to find an expression which is finite for all on-shell deformations which preserve the FG form of the metric.  A mathematically very similar problem was considered by Comp\`{e}re and Marolf
in \cite{Compere:2008us},   who defined a finite symplectic form for the case of Neumann, rather than the usual Dirichlet, boundary conditions on the metric in anti-de Sitter space.

In the usual Einstein-Hilbert story, one begins with the variation of the four-form Lagrangian
\be
\delta {L}_{EH}\left[g\right] =  d {\Theta}_{EH}\left[ g,\delta g \right] 
\ee
where the so-called presymplectic  three-form ${\Theta}_{EH}\left[ g,\delta g \right]$ is the remaining boundary term.  An expression for ${\Theta}_{EH}$ can be found in (\ref{thetaEH}). From this presymplectic three-form we can define the symplectic three-form ${\omega}_{EH}$  by
\be
{\omega}_{EH}[\delta_1 g, \delta_2 g] = \delta_1 {\Theta}_{EH} \left[g,\delta_2 g\right] - \delta_2 {\Theta}_{EH} \left[g,\delta_1 g\right]~.
\ee
The symplectic product associated to $\Sigma$ is then
\be\label{sprod}
\langle \delta_1g|\delta_2 g \rangle_{EH,\Sigma}=\int_{\Sigma}{\omega}_{EH}[\delta_1 g, \delta_2 g]~.
\ee
From the explicit expression in the appendix \ref{symplectic}  for ${\omega}_{EH}$, one may readily verify that this expression however is not in general finite for on-shell variations respecting our boundary conditions (\ref{FG}). To remedy this we note that $ \Theta_{EH}$ is ambiguous up to the addition of an exact form $d{B}$. Finiteness can be restored by a judicious choice of ${B}$, which is in fact canonically associated to the boundary counterterms added in the previous section to ensure finiteness of the Brown-York stress tensor. The result is \cite{Compere:2008us}
\be\label{omegamod}
{\omega}_{mod} \left[\delta_1 g,\delta_2 g \right] = {\omega}_{EH} \left[ \delta_1 g,\delta_2 g \right] + d {\omega}_{ct}\left[ \delta_1 \gamma,\delta_2 \gamma \right],
\ee
where  $\gamma$ is the induced metric on the boundary of $\Sigma$ and ${\omega}_{ct}$ is the symplectic form built out of the counterterm action. Explicit expressions are given in appendix \ref{symplectic}.  Using these expressions a computation essentially identical -- up to a few sign changes -- to the one in \cite{Compere:2008us} shows that the associated symplectic products are finite. Note however that local conservation of the symplectic form (\ref{omegamod}) does not imply conservation of the symplectic product (\ref{sprod}) since physical excitations leak through $\mathcal{I}^+$.

\subsection{Integrable charges}

Given a finite symplectic product, and an allowed diffeomorphism $\xi$ of the form (\ref{gdf}), the covariant phase space formalism provides a canonical construction of the charge difference between two solutions which differ by an amount $\delta g$.\footnote{For some details of the derivation we refer the reader to appendix \ref{deriv} and \cite{Compere:2008us}.}
Define
\be\label{kds}
{k}^{dS}_\xi[\delta g] =  I_\xi {\omega}_{mod} \left[ \mathcal{L}_\xi g, \delta g \right]~,
\ee
where the object $I_\xi$ is a homotopy operator whose explicit form can be found in \cite{Barnich:2001jy,Compere:2007az}. The infinitesimal charge difference for a two-dimensional compact hypersurface   $\partial\Sigma$ in $\mathcal{I}^+$ is then \be
\delta Q_\xi = \int_{\partial \Sigma} {k}^{dS}_\xi[\delta g] = \delta Q^{BY}_\xi - \frac{1}{2} \int_{\partial \Sigma} d^2 x \sqrt{g^{(0)}} n^k [g^{(0)}] \xi_k T^{ij}_{(0)} \delta g^{(0)}_{ij}~,
\ee
For future convenience, we introduce the three-form ${\Theta}_{(0)}$ through:
\be\label{theta0}
\int_{\partial \Sigma} i_\xi {\Theta}_{(0)}[g^{(0)},\delta g^{(0)}] \equiv \frac{1}{2} \int_{\partial \Sigma} d^2 x \sqrt{g^{(0)}} n^k [g^{(0)}] \xi_k T^{ij}_{(0)} \delta g^{(0)}_{ij}~,
\ee
where $i_\xi$ is the interior product with respect to $\xi^i$.
Notice that upon imposing Dirichlet boundary conditions, i.e. that the variation of the boundary metric $g^{(0)}$ vanishes, the charges are equivalent to those of the Brown-York formalism. However such a boundary condition is inappropriate at the boundary of \dsp\ because it precludes for example gravity waves.

A  second, related,  issue concerning these charge differences  is that they are not integrable: there is no unambiguous  way to integrate the infinitesimal charge differences up to a finite one. The obstruction is visible in the non-vanishing commutator
\be\label{nonint}
\left(\delta_1 \delta_2 - \delta_2 \delta_1 \right)Q_\xi =  -\delta_1 \int_{\partial \Sigma} i_\xi {\Theta}_{(0)}[g^{(0)},\delta_2 g^{(0)}] -(1 \leftrightarrow 2) \neq 0~.
\ee
Exactly the same problem is encountered in defining BMS type charges at \ip\ in Minkowski space. In this context  Wald and Zoupas \cite{Wald:1999wa} have proposed adding an additional boundary term to the charges
\be\label{waldzoupas}
\delta Q^{WZ}_\xi = \delta Q_\xi + \int_{\partial \Sigma} i_\xi {\Theta}^{WZ}\left[ g^{(0)},\delta g^{(0)} \right]~.
\ee
The ${\Theta}^{WZ}$ boundary term is designed to precisely cancel the term leading to non-integrability in (\ref{nonint}). We can immediately identify, exactly as in \cite{Wald:1999wa}:
\be
{\Theta}^{WZ} = {\Theta}_{(0)}~.\ee 
As discussed in \cite{Wald:1999wa}, the $\Theta^{WZ}$ term is related to the flux through:
\be
{F}_\xi = {\Theta}^{WZ}\left[g^{(0)}, \delta_\xi g^{(0)} \right]~,
\ee
in agreement with our earlier expression (\ref{fluxI}).


Adding this term  to the covariant phase space charge gives finally
\be\label{wz}
\delta Q_\xi^{WZ} =\delta Q_\xi^{BY} \quad \Longrightarrow  \quad Q_\xi^{WZ} = \int_{\partial\Sigma} d^2x \sqrt{\sigma} n^i \xi^j T_{ij}~.
\ee
Hence, the covariant phase space charges, after a lengthy analysis, reduce precisely to those of Brown and York. However it can now be seen explicitly that the conformal slicing transformations are indeed trivial, as earlier anticipated.

\section*{Acknowledgements}

It has been a great pleasure discussing this work with G. Comp\`ere and D. Marolf. This work was supported in part by DOE grant DE-FG02-91ER40654.

\appendix

%
%

\section{The Symplectic Form}\label{symplectic}

The symplectic form for the Einstein-Hilbert Lagrangian is given by:\footnote{For asymptotically de Sitter spacetimes obeying the Fefferman-Graham form one can show that the Iyer-Wald \cite{Wald:1999wa} and Barnich-Brandt constructions \cite{Barnich:2001jy} of the symplectic form are equivalent.}
\be
{\omega}_{EH}^\mu \left[ \delta_1 g, \delta_2 g \right] = - P^{\mu\nu\beta\gamma\epsilon\zeta} \left( \delta_2 g_{\beta\gamma} \nabla_\nu \delta_1 g_{\epsilon \zeta} - ( 1 \leftrightarrow 2 ) \right)~,
\ee
where the tensor density $P^{\mu\nu\beta\gamma\epsilon\zeta}$ is given by:
\be
P^{\mu\nu\alpha\beta\gamma\delta} = \frac{\partial}{\partial g_{\gamma\delta,\alpha \beta  }}\frac{\delta \mathcal{L}_{EH}}{\delta g_{\mu\nu}}~,
\ee
where $\mathcal{L}_{EH} \equiv \sqrt{-g}R[g]/16\pi G$. More explicitly, $P^{\mu\nu\alpha\beta\gamma\delta}$ is given by:
\begin{multline}
P^{\mu\nu\alpha\beta\gamma\delta}  = \frac{\sqrt{-g}}{32\pi G}\left( g^{\mu\nu}g^{\gamma(\alpha}g^{\beta)\delta} + g^{\mu(\gamma}g^{\delta)\nu} g^{\alpha\beta} + g^{\mu(\alpha}g^{\beta)\nu}g^{\gamma\delta}  \right. \\ \left. - g^{\mu\nu}g^{\gamma\delta} g^{\alpha\beta}  - g^{\mu(\gamma} g^{\delta)(\alpha}g^{\beta)\nu} - g^{\mu(\alpha} g^{\beta)(\gamma}g^{\delta)\nu} \right)~.
\end{multline}
The symplectic form for the counterterm action $S_{ct}$ in four-dimensions can be expressed in terms of the three-dimensional symplectic form of the Einstein-Hilbert Lagrangian as follows:
\be
{\omega}_{ct}\left[ \delta_1 \gamma, \delta_2 \gamma \right] = {\omega}^{(3)}_{EH}\left[ \delta_1 \gamma, \delta_2 \gamma \right]~.
\ee
In the FG gauge, variations of the metric reduce to variations of the induced metric $\gamma$ at the boundary and the pullback of the $(d+1)$-dimensional sympectic structure ${\omega}_{EH}$ to the spatial slice $\Sigma$ can be expressed in terms of the $d$-dimensional symplectic structure ${\omega}^{(d)}_{EH}$ as follows:
\be
{\omega}_{EH} \left[ \delta_1 g, \delta_2 g \right] = \frac{1}{\eta} {\omega}^{(d)}_{EH} \left[ \delta_1 \gamma, \delta_2 \gamma \right]~.
\ee
Expanding ${\omega}^{(d)}_{EH}$ on-shell using the FG expansion (\ref{FG}):
\begin{multline}
{\omega}^{(d)}_{EH} \left[ \delta_1 \gamma, \delta_2 \gamma \right] = \eta^{2-d} {\omega}_{(0)EH}  \left[ \delta_1 g^{(0)}, \delta_2 g^{(0)} \right] + \\  \eta^{4-d} {\omega}_{(2)EH}  \left[ \delta_1 g^{(0)}, \delta_2 g^{(0)} \right] + \mathcal{O}\left(\eta^{5-d}\right)~.
\end{multline}

\section{Covariant Phase Space Charges}\label{deriv}

In this appendix we provide some details on the charges. We begin with the expression for the charge:
\be
\delta Q_\xi \equiv \int_{\partial \Sigma} {k}^{dS}_\xi [\delta g]~,
\ee
where ${k}^{dS}_\xi [\delta g]$ was given in (\ref{kds}). In direct analogy to \cite{Compere:2008us},
we find
\be
\delta Q_\xi = -\delta \int_{\partial \Sigma} {K}_\xi\left[g\right]  + \int_{\partial \Sigma} i_\xi {\Theta}_{EH}[g,\delta g]+\int_{\partial \Sigma} {\omega}_{ct}[{\cal L}_{\xi} g, \delta g]~,
\ee
where $\partial \Sigma$ is a compact two-dimensional submanifold of $\mathcal{I}^+$ and
\be\label{thetaEH}
{\Theta}_{EH}\left[g,\delta g \right] \equiv \frac{\sqrt{-g}}{96\pi G}\left[ g^{\mu\nu} \nabla^\beta \delta g_{\beta\nu} - g^{\alpha\beta}\nabla^\mu \delta g_{\alpha\beta} \right] \epsilon_{\mu\lambda\rho\sigma}dx^\lambda dx^\rho dx^\sigma~.
\ee
The ${K}_\xi[g]$ term is the usual Komar term \cite{Komar}:
\be
{K}_\xi[g] \equiv - \frac{\sqrt{-g}}{64\pi G} \left[ \nabla^\mu \xi^\nu - \nabla^\nu \xi^\mu \right]  \epsilon_{\mu\nu\beta\gamma}dx^\beta dx^\gamma~.
\ee
The counter terms are related to $\Theta_{EH}$ by the following condition:
\be
{\Theta}_{EH} [g,\delta g]|_{\mathcal{I}^+}  = \delta {L}_{GH}+\delta \gamma_{ij} \frac{\delta {L}_{ct}}{\delta \gamma_{ij}} - \half \sqrt{g^{(0)}} T^{ij}_{(0)} \delta g^{(0)}_{ij} d^3 x~.
\ee
We further define the pre-symplectic form ${\Theta}_{ct}[\gamma,\delta\gamma]$ which is related to the counterterm sympectic structure ${\omega}_{ct}$ and counterterm Lagrangian as:
\be
d{\Theta}_{ct} \equiv \delta {L}_{ct} - \delta  \gamma_{ij} \frac{\delta {L}_{ct}}{\delta \gamma_{ij}}~,  \quad {\omega}_{ct}[\delta_1 \gamma,\delta_2\gamma] = \delta_1 {\Theta}_{ct} [\gamma,\delta_2\gamma] - \delta_2 {\Theta}_{ct} [\gamma,\delta_1\gamma]~.
\ee
Combining the above we obtain:
\be\label{charges}
\delta Q_\xi = \delta \int_{\delta\Sigma} \left( - {K}_\xi +i_\xi {L}_{GH}+ i_\xi {L}_{ct}- {\Theta}_{ct}[\gamma,\delta_\xi \gamma] \right) - \int_{\partial \Sigma} i_\xi {\Theta}_{(0)} [g^{(0)},\delta g^{(0)}]~.
\ee
It then follows from a straightforward calculation that (\ref{charges}) leads to
\be\label{finalcharge}
\delta Q_\xi = \delta Q_{BY} - \int_{\partial \Sigma} i_\xi {\Theta}_{(0)} [g^{(0)},\delta g^{(0)}]~.
\ee

\end{document}